# On the Possibility of Quantum Informational Structural Realism*

***Terrell Ward Bynum***
**Southern Connecticut State University**

**Abstract:** In *The Philosophy of Information,* Luciano Floridi presents an ontological theory of Being *qua* Being, which he calls "Informational Structural Realism", a theory which applies, he says, to every possible world. He identifies primordial information ("*dedomena")* as the foundation of any structure in any possible world. The present essay examines Floridi's defense of that theory, as well as his refutation of "Digital Ontology" (which some people might confuse with his own). Then, using Floridi's ontology as a starting point, the present essay *adds quantum features to dedomena,* yielding an ontological theory for our own universe, *Quantum Informational Structural Realism,* which provides a metaphysical interpretation of key quantum phenomena, and diminishes the "weirdness" or "spookiness" of quantum mechanics.

***Key Words***: digital ontology, dedomena, structural realism, quantum information, primordial qubit

## 1. *Introduction: Physics and the Information Revolution*

It is a commonplace today to hear people say that we are "living in the Age of Information" and that an "Information Revolution" is sweeping across the globe, changing everything from banking to warfare, medicine to education, entertainment to government, and on and on. But how can information technology (IT) enable us to transform our world so quickly and so fundamentally? Recent developments in physics, especially in quantum theory and cosmology, may provide an answer. During the past two decades, many physicists have come to believe that our universe is made of information; that is, *the universe is a vast "sea" of quantum information ("qubits"), and all objects and processes in our world (including human beings) are constantly changing quantum data structures dynamically interacting with each other.* (See, for example, Lloyd 2006 and Vedral 2010.) If everything in the world is made of information, and IT provides knowledge and tools for analyzing and manipulating information, then we have an impressive explanation of the transformative power of IT *based upon the fundamental nature of the universe.*

This essay explores some implications of that view, beginning with a discussion of key ideas from Luciano Floridi, whose "Philosophy of Information" project recently led him to a metaphysical account of the ultimate nature of the universe. That account, which he calls *Informational Structural Realism*, is similar in some ways to contemporary cosmology. Indeed, in 2008-2009 Floridi was the first philosopher, ever, to hold the prestigious post of Gauss Professor at the Göttingen Academy of Sciences in Germany (previous Gauss Professors were physicists or mathematicians). In Section 2 below, Floridi's Philosophy of Information project is briefly described in order to provide a context in which to discuss (1) his refutation of digital ontology in Section 3 and (2) his defense of Informational Structural Realism in Section 4.

After that, the remaining sections of this essay explore the possibility of a *quantum variation* of Floridi's Informational Structural Realism, a variation which reinterprets Floridi's

"*Ur*-relation" (2011, p. 354) to be quantum in nature. As explained below, Floridi's Informational Structural Realism applies to all possible worlds, but I wish to focus here only upon *this* world. One result of this strategy is a metaphysical interpretation of quantum mechanics, which is briefly described in Sections 5 to 9 below.

## 2. *Floridi's Philosophy of Information Project*

In the late 1990s, Floridi launched a project that he called "The Philosophy of Information". His ambitious goal was to create a paradigm that would someday become part of the "bedrock" of philosophy (*philosophia prima*). The central concept was to be *information*, a concept with multiple meanings, and also, according to Floridi,

> a concept as fundamental and important as being, knowledge, life, intelligence, meaning, or good and evil – all pivotal concepts with which it is interdependent – and so equally worthy of autonomous investigation. It is also a more impoverished concept, in terms of which the others can be expressed and interrelated, when not defined. (Floridi 2002, p. 134)

Floridi's philosophical method is that of *constructionism*, which holds that ultimate reality (the "noumenal" world of "things-in-themselves", as Kant would say) is essentially unknowable — except, according to Floridi, the fundamental ontological claim of Informational Structural Realism, which he derives using a kind of transcendental argument (see Section 4 below). Ultimate reality, says Floridi, provides certain affordances and imposes certain constraints upon our experiences, observations, and experiments, but we are forever unable to know how and why it does so. The best that we can do is to construct models of reality, or parts thereof. Knowledge, truth and semantics apply to our models, and not to ultimate reality itself because we cannot know *that*. The world in which we live (Kant's "phenomenal" world) is the sum total of our models of reality, so if we significantly change the objects and/or processes within our models, then we live in a different phenomenal world. It is important to note, however, that this is not a version of relativism, because models can be compared with regard to their ability to accommodate the constraints and affordances of the unknowable ultimate reality.

Floridi constructs his models using a "method of abstraction" that he and his colleague J. W. Sanders adapted from Formal Methods in computer science. When using the Floridi-Sanders method of abstraction, one selects a set of "observables" at a given "level of abstraction"; then, by attributing certain "behaviors" to the observables, one builds a model of the entity being analyzed. Finally, the resulting model is tested against experiences, observations and experiments. The best models are those that most successfully achieve "informativeness, coherence, elegance, explanatory power, consistency, predictive power, etc." (Floridi 2011, p.348)

During the past decade, Floridi's Philosophy of Information project has become a broad research program addressing a wide array of philosophical questions. These range from the deceptively simple question, "What is information?", to topics such as the nature and ethics of artificial agents, the foundation and uniqueness of computer ethics, the semantics of scientific models, the nature and role of artificial companions in a human life, the informational nature of the universe, symbol grounding and consciousness, the role of information in reasoning and logic, and many more. (See, for example, his book *The Philosophy of Information*, 2011.) The next two sections of the present essay concern Floridi's account of the informational nature of the universe, beginning, in Section 3, with his refutation of a view that is often called "digital ontology".

### 3. *Floridi's Refutation of Digital Ontology*

In Chapter 14 of *The Philosophy of Information*, Floridi carefully distinguishes between his own account of the ultimate nature of the universe — Informational Structural Realism — and digital ontology, a theory that some people might, *mistakenly*, assume to be his. Floridi's refutation focuses especially upon versions of digital ontology that presuppose the Zuse Thesis advocated by the German computer scientist Konrad Zuse:

> The universe is being deterministically computed on some sort of giant but discrete computer. (Zuse 1967, 1969) [quoted in Floridi 2011, p. 319]

Such theories are summarized by Edward Fredkin as theories which are

> based upon two concepts: bits, like the binary digits in a computer, correspond to the most microscopic representation of state information; and the temporal evolution of state is a digital informational process similar to what goes on in the circuitry of a computer processor. (Fredkin 2003, p. 189) [quoted in Floridi 2011, p.318]

To begin his refutation of this kind of digital ontology, Floridi provides the following summary:

> The overall perspective, emerging from digital ontology, is one of a metaphysical monism: ultimately, the physical universe is a gigantic digital computer. It is fundamentally composed of digits, instead of matter or energy, with material objects as a complex secondary manifestation, while dynamic processes are some kind of computational states transitions. There are no digitally irreducible infinities, infinitesimals, continuities, or locally determined random variables. In short, the ultimate nature of reality is not smooth and random but grainy and deterministic. (Floridi 2011, p. 319)

Most versions of digital ontology typically presuppose that the entire physical universe is an enormous computer (pancomputationalism). Nevertheless, digital ontology *can* be separated from pancomputationalism; and, indeed, some versions of pancomputationalism (for example, Laplace's) are analogue rather than digital. Floridi makes it clear that his Informational Structural Realism is *neither digital nor analogue*; and he also notes that it is *not* committed, one way or the other, to pancomputationalism.

It is my view that Floridi's case against digital ontology, as defined above, is strong. Readers interested in the step-by-step details are referred to Chapter 14 of *The Philosophy of Information*. Here, I want to summarize some of Floridi's key points against digital ontology in order to set the stage for discussions below about his Informational Structural Realism, and about my suggested quantum variant of it, which is not subject to any of the objections listed here:

*Criticism (i), Digital Ontology Requires More Digital Memory than Is Possible*: If one assumes (like Fredkin, quoted above, for example) that ultimate reality consists of classical bits being processed like those in a traditional computer, our current scientific understanding of the universe would lead us to conclude that *the evolution of the universe since the Big Bang could not have occurred* because there would not have been enough digital memory. Floridi explains (2011, p. 323):

> Here is a very simple illustration: Lloyd (2002) estimates that the physical universe, understood as a computational system, could have performed $10^{120}$ operations on $10^{90}$ bits [. . . ] since the Big Bang. The problem is that if this were true, the universe would 'run out of memory':
>
> > To simulate the Universe in every detail since time began, the computer would have to have $10^{90}$ bits — binary digits, or devices capable of storing a 1 or a 0 — and it would have to perform $10^{120}$ manipulations of those bits. Unfortunately, there are probably only around $10^{80}$ elementary particles in the Universe. (Ball (2002, 3 June)) [quoted in Floridi 2011, p. 323]

It is important to note that the "bits" of digital ontology, as defined here, are *traditional binary bits* that can be *either 1 or 0 but not both*. Therefore, criticism (i) would *not* apply to quantum bits (qubits)*,* which can be *both* 1 and 0 at the same time, as well as an infinite number of states between 1 and 0 (see Section 5 below).

*Criticism (ii), Digital Ontology Requires a Radical Change in Current Scientific Practice*: A second criticism of digital ontology (Floridi 2011, p. 324) is the fact that "its success would represent a profound change in our scientific practices and outlook". Since a significant amount of current science is based upon powerful analogue ideas like force fields, waves, continuous functions, differential equations, Fourier transforms, and so on, this places a heavy burden of proof upon advocates of digital ontology, who would have to show that the powerful analogue ideas of contemporary science can be successfully reinterpreted digitally.

*Criticism (iii), Digital Ontology Misapplies the Concepts "Digital" and "Analogue"*: Even if defenders of digital ontology could — somehow — reinterpret the powerful analogue concepts of contemporary science, Floridi argues that "it is not so much that reality in itself is not digital, but rather that, in a metaphysical context, the digital vs analogue dichotomy is not applicable." He introduces a thought experiment to demonstrate that the concepts "digital" and "analogue" apply only within our models of reality. They are features of our models "adopted to analyze reality, not features of reality in itself." Some models are analogue and some are digital, and we are unable to know whether reality itself is either of these or something to which neither concept can be applied. To overcome this impasse, Floridi "seeks to reconcile digital and analogue ontology by identifying the minimal denominator shared by both." Thus, Floridi adopts the following strategy:

> What remains invariant [in ultimate reality, given our model-building perspective] cannot be its digital or its analogue nature, but rather the structural properties that give rise to a digital or analogue reality. These invariant, structural properties are those in which science is mainly interested. So it seems reasonable to move from an ontology of things — to which it is difficult not to apply the digital/discrete vs analogue/continuous alternative — to an ontology

> of structural relations, to which it is immediately obvious that the previous dichotomy is irrelevant. (2011, p. 334)

Floridi's case against digital ontology is intended to clear the way for his positive defense of Informational Structural Realism in Chapter 15 of *The Philosophy of Information*. His move "to an ontology of structural relations" is central to that defense, which is discussed in the next section.

### 4. *Floridi's Informational Structural Realism*

In presenting his positive case for Informational Structural Realism, Floridi agrees with Putnam's "No-Miracles Argument":

> (Some form of) realism 'is the only philosophy that does not make [the predictive success of] science a miracle' (Putnam 1975, p. 73) [quoted by Floridi on p. 345]

Like every version of realism, Floridi's presupposes that there exists "a mind-independent reality addressed by, and constraining, knowledge". In addition, his theory supports the adoption of models which "carry a minimal ontological commitment in favour of the structural properties of reality and a reflective, equally minimal, ontological commitment in favour of structural objects." (2011, p. 339) Unlike other versions of structural realism, though, Floridi's theory

> supports *an informational interpretation of these structural objects.*This second commitment [ . . . ] is justified by epistemic reasons. We are allowed to commit ourselves ontologically to whatever minimal conception of objects is useful to make sense of our first commitment in favour of structures. The first commitment answers the question 'what can we know?'; and the second commitment answers the question 'what can we justifiably assume to be in the external world?'. (2011, p. 339) [my emphasis added here]

The "structural objects" that Floridi presupposes — the primordial "*Ur*-relations" of the universe — are what he calls *dedomena*: "mind-independent points of lack of uniformity in the fabric of Being" — "mere *differentiae de re"* (he also refers to them, metaphorically, as "data in the wild"). These cannot be directly perceived, and they cannot be detected by any kind of scientific instrument. Instead, Floridi infers their existence by a transcendental argument according to which *dedomena must exist to make it possible for any structured entities at all to exist*.

> Dedomena are not to be confused with environmental data. They are pure data or proto-epistemic data, that is, data before they are epistemically interpreted. As 'fractures in the fabric of Being', they can only be posited as an external anchor of our information, for dedomena are never accessed or elaborated independently of [an epistemic model of reality]. They can be reconstructed as ontological requirements, like Kant's *noumena* or Locke's *substance*: they are not epistemically experienced, but their presence is empirically inferred from, and required by, experience. Of course, no example can be provided, but dedomena are whatever lack of uniformity in the world is the source of (what looks to informational organisms like us as) data [. . . ]. (2011, Ch. 4, pp. 85-86)

Floridi makes a case for the view that the ultimate nature of any possible universe must include at least some dedomena, because the relation of difference is a precondition for any other relation:

> Let us consider what a completely undifferentiable entity *x* might be. It would be one unobservable and unidentifiable at any possible [level of abstraction]. Modally, this means that there would be no possible world in which *x* would exist. And this simply means that there is no such *x*. [ . . . ] Imagine a toy universe constituted by a two-dimensional, boundless, white surface. Anything like this toy universe is a paradoxical fiction that only a sloppy use of logic can generate. For example, where is the observer in this universe? Would the toy universe include (at least distinguishable) points? Would there be distances between these points? The answers should be in the negative, for this is a universe without relations. (2011, Ch. 15, p. 354)

Thus, there can be no possible universe without relations; and since dedomena are preconditions for *any* relations, it follows that every possible universe must be made of at least some dedomena. (Note that there might also be other things which, for us, are forever unknowable.) There is much more to Floridi's defense of Informational Structural Realism, including his replies to ten possible objections, and I leave it to interested readers to find the details in Chapter 15 of *The Philosophy of Information*.

In the present essay, I assume that Floridi has made his case for dedomena as components in the "underlying fabric" of every possible world, including our own. He views the fact that his ontology applies to every possible world as a very positive feature. It means, for example, that Informational Structural Realism has maximum "portability", "scalability", and "interoperability".

Regarding *portability*, Floridi notes that:

> The most portable ontology would be one that could be made to 'run' in any possible world. This is what Aristotle meant by a general metaphysics of Being *qua* Being. The portability of an ontology is a function of its importability and exportability between theories even when they are disjointed ([their models] have no observables in common). Imagine an ontology that successfully accounts for the natural numbers and for natural kinds. (p. 357)

*Scalability*, according to Floridi, is the capacity of a theory to work well even when "the complexity or magnitude of the problem increases."

> Imagine an ontology that successfully accounts not only for Schrödinger's cat but also for the atomic particles dangerously decaying in its proximity. (p. 357)

The *interoperability* of an ontology is "a function of its capacity of allowing interactions between different [scientific or common-sense] theories." Floridi illustrates this by inviting us to

> Imagine an ontology that successfully accounts for a system modeled as a brain and as a mind. (p. 358)

Using these three notions — portability, scalability, and interoperability — Floridi introduces the concept of “a specific metaphysics”, which he defines as “an ontology with fixed degrees of portability, scalability, and interoperability” (p. 358). It is possible to criticize a specific metaphysics if it is “too local”, in the sense that its degrees of portability or scalability or interoperability are limited. Thus, he notes:

> For example, a Cartesian metaphysics is notoriously undermined by its poor degree of interoperability: the mind/body dualism generates a mechanistic physics and a non-materialist philosophy of mind that do not interact very well. Leibniz’s metaphysics of monads is not easily scalable (it is hard to account for physical macro-objects in its terms). (p. 358)

The most “local” kind of ontology would be naïve realism, because it assumes that a model is a direct and accurate representation of the modeled entity. At a given moment in the history of science, this could make naïve realism appear to be very strong; but, as Floridi points out, it is “dreadfully brittle” and “easily shattered by any epistemic change”, even by a simple counter example or by a skeptical argument, rather than a whole scientific revolution.

In my view, Floridi has successfully argued for Informational Structural Realism, including his transcendentally inferred assumption that every possible world must include dedomena within its underlying fabric of reality. As explained in the next section, however, I also believe — and I think that Floridi would agree — that metaphysical theories which do not apply to every possible world can nevertheless be philosophically rewarding and worthy of consideration in appropriate circumstances.

As an example of a “more local” metaphysics, which nevertheless is worthy of one's consideration, I suggest adding quantum properties to Floridi’s dedomena to generate an ontology that would apply to our own world (and any other world that happens to include quantum structures). Such a metaphysics would not attempt to explain Being *qua* Being, like Aristotle's or Floridi's; but perhaps it could aid our philosophical understanding — and maybe even our scientific understanding — of *this particular world*. In the remaining sections of this essay, therefore, I will explore the idea of trying to develop a *quantum variant* of Floridi’s Informational Structural Realism.

### 5. *The Possibility of Quantum Informational Structural Realism*

To begin a metaphysical thought experiment, let us adopt an epistemological justification modeled upon that of Floridi (see Section 4 above):

> For epistemic reasons, we are allowed to commit ourselves ontologically to whatever minimal conception of objects is useful to make sense of our first commitment in favor of *quantum structures*. The first commitment answers the question ‘what can we know?’; and the second commitment answers the question ‘what can we justifiably assume to be in the external world, *given the existence of quantum structures*?’ [my changes are in italics]

In Floridi’s case, the required primordial entities are “dedomena” — “mind-independent points of lack of uniformity"— "mere *differentiae de re*" — primordial data. These must be part of the ultimate fabric of any world (including our own) where at least one structure, no matter how minimal, exists. In *our* case, the primordial data that we need must account for the existence of quantum information — *qubits* — physical entities in our universe which can represent, simultaneously, *0 and 1 and an infinite set of numbers between 0 and 1*. Prerequisites of such entities would be mind-independent dedomena-"packets" containing an infinite number of dedomena for each qubit in our universe. If we assume the existence of such "packets" — let us call them "primordial qubits" (PQs) or "primordial quantum data" — we can provide an oportunity for creative philosophers to develop metaphysical explanations of quantum phenomena and, perhaps, even eliminate some of the alleged "weirdness" or "spookiness" of such phenomena (see below). In the spirit of this challenge, I briefly summarize, in the following sections, several "weird" phenomena of quantum mechanics, and I attempt to cast some philosophical light upon them.

The key move in the present thought experiment is to "think outside of the box" — or, as I prefer to say, *think outside of the "quantum-foam bubble" which is our universe* (see below). Imagine, for want of a better metaphor, a vast primordial PQ "ocean" or source. Conceivably, such a source could contain many other things besides PQs; but, for our purposes, we need only assume that the primordial "ocean" is a vast source of PQs. Given this assumption, the birth of our universe (the Big Bang) can be interpreted as the sudden appearance of a constantly expanding "bubble" (see below) immersed in the primordial PQ ocean. Instead of air, the bubble is filled with *quantum foam*, a "medium" which consists of an enormous number of "virtual quantum particles":

*Quantum Foam*: In our universe, totally empty space does not exist. Thus, even if all of the usual matter and electromagnetic radiation were removed from a given region of outer space, leaving only what is sometimes called "the quantum vacuum", there would remain what physicist John Wheeler called "quantum foam" — "virtual quantum particles" that are constantly coming into existence, interacting with each other, and disappearing within a tiny fraction of a second. As physicist Frank Close explains, in his book *Nothing: A Very Short Introduction*, "When viewed at atomic scales, the Void is seething with activity, energy and particles." (Close 2009, p. 94) He went on to note, later in that same book, that

> There is general agreement [among physicists] that the quantum vacuum is where everything that we know came from, even the matrix of space and time. . . . the seething vacuum offers profound implications for comprehending the nature of Creation from the Void [i.e., creation from quantum foam]. (p. 106)

And also,

> the multitude of disparate phenomena that occur at macroscopic distances, such as our daily experiences, are controlled by the quantum vacuum [i.e., the quantum foam] within which we exist. (p. 122)

Given these ideas from contemporary physics, the present metaphysical thought experiment yields the following account of the birth and nature of our universe:

> In the beginning was the primordial qubit source (the "PQ ocean"). The birth of our

> universe (the Big Bang) was the formation and very rapid expansion of a quantum-foam bubble (our universe) within the PQ ocean. Initially, the PQs in the ocean interacted with the bubble very rapidly, generating additional quantum foam and an explosive expansion of the bubble (called "inflation" by physicists). During the Big Bang, quantum laws together with quantum foam generated elementary particles and the spacetime matrix. As the rapidly expanding bubble began to cool, the various kinds of "standard-model" quantum particles came into existence, including — eventually — the Higgs boson. With the arrival of the Higgs boson, the rate of expansion dramatically decreased but was not entirely eliminated. Our universe continues to expand at an accelerating rate as the PQ ocean generates more and more quantum foam within it. (Perhaps the increasing quantum foam is the "dark energy" that is accelerating the expansion of our universe.)

A key assumption of this metaphysical thought experiment is that quantum phenomena, such as *superpositions, decoherence, entanglement, "spooky action at a distance",* and *teleportation* (see below), should be viewed, not as weird and inexplicable phenomena, but rather as scientific evidence that casts light upon the nature of the primordial quantum data source and upon quantum foam. In the sections below, this metaphysical "model" is used to interpret several important quantum phenomena.

### 6. *"It from bit" — To be is to be a quantum data structure*

In 1990, in an influential paper, physicist John Wheeler introduced his famous phrase "it from bit" (Wheeler 1990), and he thereby gave a major impetus to an information revolution in physics . In that paper, Wheeler declared that "all things physical are information theoretic in origin" — that "every physical entity, every it, derives from bits" — that "every particle, every field of force, even the spacetime continuum itself . . . derives its function, its meaning, its very existence" from bits. He predicted that "Tomorrow we will have learned to understand and express *all* of physics in the language of information." (emphasis in the original)

Since 1990, a number of physicists — some of them inspired by Wheeler — have made great strides toward fulfilling his "it-from-bit" prediction. In 2006, for example, in his book *Programming the Universe*, Seth Lloyd presented impressive evidence supporting the view that the universe is not only a vast sea of qubits, it is actually a gigantic quantum computer:

> The conventional view is that the universe is nothing but elementary particles. That is true, but it is equally true that the universe is nothing but bits — or rather, nothing but qubits. Mindful that if it walks like a duck and it quacks like a duck then it's a duck . . . since the universe registers and processes information like a quantum computer, and is observationally indistinguishable from a quantum computer, then it *is* a quantum computer. (p. 154, emphasis in the original)

More recently, in 2011, three physicists used axioms from information processing to derive the mathematical framework of quantum mechanics (Chiribella, et al. 2011) . These are only two of a growing number of achievements that have begun to fulfill Wheeler's "it from bit" prediction.

If, for purposes of the present essay, we assume that the "bits" which Wheeler mentioned in his "it from bit" prediction, are qubits, then Wheeler's view would be that qubits are responsible for the very existence of every particle and every field of force — even for spacetime itself. This, in turn, would mean that qubits must have existed prior to every other thing in our universe, and they must have been involved in the Big Bang. As Seth Lloyd has said, "*The Big Bang was also a Bit Bang*" (Lloyd 2006, p. 46); and he noted elsewhere that the motto of his own understanding of the nature of the universe is "It from qubit". (Lloyd 2006, p. 175)

Objects comprised of qubits exhibit quantum features like *genuine randomness, superposition* and *entanglement* – features that Einstein and other scientists considered "weird" and even "spooky". As explained below, these scientifically verified quantum phenomena raise important questions about traditional bedrock philosophical concepts. If every physical thing in the universe consists of qubits, then one would expect that *any* physical entity could be in many different states at once, depending upon the many states of the qubits of which it is composed. Indeed, some quantum physicists have noted that, under the right circumstances, "All objects in the universe are capable of being in all possible states" (Vedral 2010, p. 122). This means that there is a scientifically verifiable sense in which objects comprised of qubits can be in many different places at once. It means that biological beings — like Schrödinger's famous cat or a human being — could be both alive and dead at the same time, and at least some things can be teleported from place to place instantly, faster than the speed of light, without passing through the space in between. Finally, it also means that, at the deepest level of reality, the universe is both digital and analogue at the same time. These are not mere speculations, but requirements of quantum mechanics, which is the most tested and most strongly confirmed scientific theory in history. It is of interest to note that, because of these scientifically confirmed facts about the world, philosophers will have to rethink many fundamental philosophical concepts, like *being* and *non-being*, *real* and *unreal*, *actual* and *potential*, *cause* and *effect*, *consistent* and *contradictory, knowledge* and *thinking*, and many more. (See below.)

### 7. *Coming into existence in the classical world*

A familiar "double-slit experiment", which is often performed today in high school physics classes and undergraduate laboratories, illustrates the ability of different kinds of objects to be in many different states at once. In such an experiment, objects are fired, one at a time, by a "gun" toward a screen designed to detect them. The objects in the experiment, can be, for example, photons, or electrons, or single atoms, or much larger objects, such as "buckeyballs" (composed of 60 carbon atoms comprised of 1,080 subatomic particles), or even larger objects.

To begin a double-slit experiment, a metal plate with two parallel vertical slits is inserted between the gun and the detection screen. The gun then fires individual objects — one at a time — toward the double-slit plate. If the objects were to act like classical ones, some of them would go through the right slit and strike the detection screen behind that slit, while others would go through the left slit and strike the detection screen behind that slit. But this is not what happens. Instead, surprisingly, *a single object goes through both slits simultaneously*, and when a sufficient number of individual objects has been fired, a *wave-interference pattern* is created on the detection screen from the individual spots where the objects randomly landed. In such an experiment, an individual object travels toward the double-slit plate *as a wave*; and then, on the other side of the double-slit plate, it travels toward the detection screen *as two waves interfering with each other*. When the two interfering waves arrive at the detection screen, however, a classical object suddenly appears on the screen *at a specific location* which could not have been known in advance, even in principle. In a double-slit experiment, then, single objects behave also like waves — even like two waves creating an interference pattern.

How is a philosopher to interpret these results? Perhaps we could try to make sense of this behavior by adopting a distinction much like Aristotle's distinction between the *potential* and the *actual*. When a child is born, for example, Aristotle would say that the child is *potentially* a language speaker, but not *actually* a language speaker. The potential of the child to speak a language is, for Aristotle, *something real* that is included in the very nature of the child. In contrast, a stone or a chunk of wood does not have the potential ever to become a language speaker. For Aristotle, *the potential and the actual are both real in the sense that both are part of the nature of a being; and the potential of a being becomes actualized through interactions with similar actualized things in the environment.* Thus a child who is *not yet* an actual language speaker, becomes an actual language speaker by interacting appropriately with people in the community who already are actual language speakers. Similarly, an unlit candle, which potentially has a flame at the top, becomes a candle with an actual flame when it interacts appropriately with some actual fire in its environment.

If we adopt a distinction that is very similar to Aristotle's, we could say that the wave in a double-slit experiment consists of *a "wave-form bundle" of possible paths* that the object could follow on its way to the detection screen. Indeed, this is an interpretation that some quantum scientists accept. The possible paths, then, are *real* entities that travel through space-time together as a wave-form bundle of physical possibilities. But where is the *actual (*that is, *classical)* object while the wave of possibilities is traveling to the screen? Has the classical object itself disappeared? Or does it exist *as* a bundle of possibilities? Typical philosophical ideas about real and unreal, cause and effect, potential and actual don't seem to fit this case. Nevertheless, double-slit experiments are regularly performed in high school classrooms and undergraduate labs around the world — and always with the same allegedly "weird" results. Quantum mechanics predicts that every object in the universe, no matter how large, would behave the same way under the right circumstances.

In quantum mechanics, the possibilities that form a quantum wave are said to be "superposed" upon each other, and so together they are called the "*superpositions*" of the quantum object. Some quantum scientists would say that the object exists *everywhere at once within the wave*. Other scientists would say that *no actual classical object exists within the wave*, and it is illegitimate even to ask for its specific location. In any case, when a wave-form bundle of possibilities interacts appropriately with another physical entity in its environment, *by sharing some information with the other physical entity*, all the "superposed" possibilities — *except one* — suddenly disappear and *one* actualized classical object instantly appears randomly at a specific location. Quantum physicists call this phenomenon, in which a wave of possibilities gets converted into an actualized classical object, *decoherence*.

Decoherence, then, is a remarkable phenomenon. It is what brings into existence actualized classical objects — located at specific places and with specific properties that can be observed and measured. Decoherence "extracts" or "creates" classical objects out of an infinite set of physical possibilities within our universe. This "extraction" process is *genuinely random*. As Anton Zeilinger explains,

> The world as it is right now in this very moment does not determine uniquely the world in a few years, in a few minutes, or even in the next second. The world is open. We can give only probabilities for individual events to happen. And it is not just our ignorance. Many people believe that this kind of randomness is limited to the microscopic world, but this is not true, as the [random] measurement result itself can have macroscopic consequences. (Zeilinger 2010, p. 265)

Random or not, *being or existing* in our universe has two different, but closely interrelated, varieties:

1. One is *quantum existence* as a wave — a bundle — of superposed physical possibilities, while the other form of existence is

2. *Classical existence* as a specific object located at a specific place in space-time with classical properties that can be observed and measured.

In our universe, the quantum realm and the classical realm exist together and constantly interact with each other. The source of classical, measurable entities is a continuously expanding array of qubits that, together, establish what is physically possible by creating an infinite set of superposed physical possibilities. From this infinite, always expanding, set of possibilities, the sharing of information generates everyday classical objects at specific locations with observable and measurable properties. This is the process of decoherence. Thus, when human beings interact with quantum entities, thereby exchanging information with them, the interaction randomly converts certain physical possibilities into actualized classical objects. In this sense, as Wheeler has said, "this is a participatory universe" in which human activities actualize classical objects (Wheeler 1990). *Quantum information, then, is the underlying source of classical physical entities. It from qubit!*

If one adopts Lloyd's view of the universe as *a quantum computer* (see Section 6 above), perhaps one could even interpret each superposition of a quantum entity as something very like a subroutine within the quantum computer, ready to be activated if and when an appropriate bit of information is received from a measurement or other physical interaction.

## 8. *Additional Quantum Phenomena*

Other quantum phenomena, such as *entanglement, "spooky action at a distance", teleportation*, and *quantum computing,* raise important questions that philosophers need to address. So, each of these phenomena is briefly discussed below along with some philosophical questions that arise from them.

*Entanglement and "Spooky Action at a Distance"* — As indicated above, a quantum entity exists as a bundle of superposed physical possibilities. An electron, for example, could exist as a bundle of superpositions which have an "up" spin and a "down" spin at the very same time. When one observes or measures the electron, which is in many different quantum states, its spin — instantly and randomly — becomes definitely "up" or definitely "down". This decoherence occurs when the electron, which had been in many superpositions, suddenly shares information with the measurer (or something else in its environment).

Sometimes two quantum entities interact in such a way that *their superpositions become "entangled" and they begin to behave like a single quantum entity*. For example, two entangled electrons each have superpositions in which their spins are up and down at the same time. Because they are entangled, however, if one electron is observed or otherwise measured, thereby randomly making its spin definitely up or definitely down, *the other electron's spin must instantly become the opposite of the spin of the first one.* The amazing thing, and some would say "puzzling" thing (Einstein said "spooky"), is that when entanglement occurs, it can continue even if the two entangled entities become separated by huge distances. Thus if one entangled electron, for example, is on earth and the other one is sent to Mars, they still can remain entangled. So, if someone measures the electron which is on earth, yielding a definite up-spin result for the earth-bound electron, then the other entangled one — the one on Mars — must instantly have a down spin! *This instant result occurs no matter how far apart the two electrons are,* and this violates the speed of light requirement of relativity theory. This is why Einstein called such an occurrence "spooky action at a distance".

Given the metaphysical model developed in this essay, the "spookiness" of entanglement can be eliminated by assuming that the entanglement consists of an interconnection *within the quantum foam medium, but outside of the spacetime matrix.* Entanglement, then, can be interpreted as something very like a *hyperlink,* outside of spacetime, that connects the superpositions of the quantum entities. When one of the entangled entities is measured, and thereby decoheres, the other entity instantly decoheres in the opposite way, without having to receive a message that travelled through spacetime.

So, given the present metaphysical thought experiment, entities in the "classical" world — including *spacetime and gravity* — are generated by the underlying quantum foam (perhaps assisted by the primordial PQ ocean). But the "laws of nature" of the classical world — such as Einstein's speed of light requirement — apply in the classical realm, while "spooky action at a distance" is generated outside of spacetime.

Another quantum phenomenon that presents a challenge to traditional philosophy is called *teleportation*, a process in which quantum properties of one object are transferred instantly to another quantum object by means of entanglement and measurement. Because the transfer of quantum properties takes place via entanglement, *it occurs outside of spacetime, no matter how far apart the two objects might be in the classical world, and without the need to travel through spacetime.* Again, given the assumptions of the present metaphysical thought experiment, the "weirdness" or "spookiness" of teleportation is diminished because Einstein's speed-of-light requirement does not apply outside of spacetime.

*Quantum Computing* — Because qubits can be simultaneously in many different states between 0 and 1, and because of the phenomenon of entanglement, quantum computers are able to perform numerous computing tasks at the very same time. As Vlatko Vedral explains,

> *any problem* in Nature can be reduced to a search for the correct answer amongst several (or a few million) incorrect answers. . . . [and] unlike a conventional computer which checks each possibility one at a time, quantum physics allows us to check multiple possibilities simultaneously. (Vedral 2010, p. 138, emphasis in the original)

Once we have learned to make quantum computers with significantly more than 19 qubits of input — which is the current state of the art — quantum computing will provide remarkable efficiency and amazing computing power. As Seth Lloyd has explained,

> A quantum computer given 10 input qubits can do 1,024 things at once. A quantum computer given 20 qubits can do 1,048,576 things at once. One with 300 qubits of input can do more things at once than there are elementary particles in the universe. (Lloyd 2006, pp. 138-139)

For philosophy, such remarkable computer power has major implications for concepts such as *knowledge*, *thinking* and *intelligence* — and, by extension, *artificial intelligence*. Imagine an artificially intelligent robot whose "brain" includes a quantum computer with 300 qubits. The "brain" of such a robot could do more things simultaneously than all the elementary particles in the universe! Compare that to the problem-solving abilities of a typical human brain. Or consider the case of so-called human "idiot savants", who can solve tremendously challenging math problems "in their heads" instantly, or remember every waking moment in their lives, or remember, via a "photographic memory", every word on every page they have ever read. Perhaps such "savants" have quantum entanglements in their brains which function like quantum computers. Perhaps *consciousness* itself is an entanglement phenomenon. The implications for epistemology and the philosophy of mind are staggering!

### 9. *Concluding Remarks*

One result of the above discussion is the conclusion that *every object and process in our universe, at the deepest level of physical existence, is a quantum data structure.* To quote Lloyd, "It from qubit!" It is a common belief that this is true only of tiny subatomic entities, and not true of larger entities; but that is incorrect. In the June 2011 issue of *Scientific American* , quantum scientist Vlatko Vedral made an impressive case (see Vedral 2011) for the view that quantum properties are not confined to tiny subatomic particles. Most people, he noted, even including a number of physicists, make the mistake of dividing the world into two kinds of entity: on the one hand, tiny particles which are quantum in nature; and on the other hand, larger "macro" objects, which obey the classical laws of physics, including relativity.

> Yet this convenient partitioning of the world is a myth. Few modern physicists think that classical physics has equal status with quantum mechanics; it is but a useful approximation of a world that is quantum at all scales. (Vedral 2011, pp. 38 and 40)

Vedral went on to discuss a number of "macro" objects which, apparently, have exhibited quantum properties, including for example, (1) entanglement within a piece of lithium fluoride made from trillions of atoms, (2) entanglement within European robins who apparently use it to guide their yearly migrations between Europe and central Africa, and (3) entanglement within plants that appear to use it to bring about photosynthesis.

Another result of the above discussion is that the metaphysical theory, which I have called *Quantum Informational Structural Realism*, provides a philosophical interpretation of quantum phenomena — including qubits, superpositions, decoherence, entanglement, teleportation and quantum computing — that is consistent with contemporary physics and diminishes the alleged "weirdness" or "spookiness" frequently attributed to quantum phenomena.

***Acknowledgements***
Research funded, in part, by grants from the Connecticut State University System and Southern Connecticut State University. I am especially grateful to the International Association for Computing and Philosophy (IACAP) for the opportunity to present an early version of this work as the Preston Covey Award Address at the IACAP2011 Conference in Aarhus, Denmark, July 2011. In addition, I benefited significantly from comments and suggestions from faculty colleagues at Southern Connecticut State University, especially Matthew Enjalran, Ken Gatzke, Krystyna Gorniak-Kocikowska, Beth Krancberg and Heidi Lockwood.